\documentclass[10pt,conference]{IEEEtran}
\IEEEoverridecommandlockouts
\usepackage{amsmath,amssymb,amsfonts}
\usepackage{algorithmic}
\usepackage[ruled]{algorithm2e}
\usepackage{array}
\usepackage{textcomp}
\usepackage{stfloats}
\usepackage{url}
\usepackage{color}
\usepackage{verbatim}
\usepackage{graphicx}
\usepackage{tabularx}
\usepackage{multirow}
\usepackage{array}
\usepackage{booktabs}
\usepackage{graphicx}
\usepackage{parskip}

\usepackage[numbers]{natbib}
\usepackage{bm}
\usepackage{mathtools}

\usepackage{subfigure}
\usepackage{booktabs}
\usepackage{multirow}
\usepackage{makecell}
\usepackage[table]{xcolor}

\usepackage{bibspacing}
\graphicspath{ {./fig/} }
\usepackage{enumitem}

\def\BibTeX{{\rm B\kern-.05em{\sc i\kern-.025em b}\kern-.08em
    T\kern-.1667em\lower.7ex\hbox{E}\kern-.125emX}}
\begin{document}

\title{TrustFormer: Cross-Temporal and Cross- Dimensional Transformer for Task-Specific Multi-Dimensional Trust Evaluation}

\author{\IEEEauthorblockN{Botao~Zhu and Xianbin~Wang}
\IEEEauthorblockA{Department of Electrical and Computer Engineering, Western University,
London, Ontario, Canada}}




\maketitle

\begin{abstract}
In dynamic collaborative systems, the selection of reliable collaborators is critical to ensuring effective task execution. Existing trust evaluation methods often rely on unidimensional or scalar representations, which fail to faithfully capture a collaborator’s true trustworthiness, thereby motivating a shift toward multi-dimensional trust modeling. However, due to the asynchrony of collected trust-related data across different dimensions, as well as the complex intra- and inter-dimensional dependencies embedded within these data, multi-dimensional trust evaluation remains challenging. To address these challenges, we propose TrustFormer, a task-specific multi-dimensional trust evaluation framework. Specifically, TrustFormer leverages task identifiers and device-generated timestamps to synchronize heterogeneous trust-related data across historical collaborations. It further employs cross-temporal and cross-dimensional attention mechanisms to jointly model temporal dynamics and inter-dimensional correlations, thereby effectively learning the multi-dimensional trust evolution of potential collaborators from historical performance data. In addition, according to the multi-dimensional resource requirements of tasks, potential collaborators' multi-dimensional resource trust is evaluated. Finally, by synthesizing these multi-dimensional trust profiles, the framework enables the optimal collaborator selection. Experimental results demonstrate that TrustFormer outperforms existing methods by yielding a 40.8\% improvement in trust evaluation accuracy and enabling more reliable collaborator selection.


\end{abstract}


\section{Introduction}
By integrating the Internet of Things (IoT), edge computing, and artificial intelligence, distributed intelligent systems are expected to support a wide range of applications, such as autonomous driving and robotic collaboration. Due to their inherent resource constraints, individual devices are unable to execute increasingly complex tasks independently, necessitating collaboration with other devices for task execution. Under this paradigm, the effective selection of reliable collaborators is of paramount importance for successful task completion and overall system effectiveness~\cite{8080202}. Trust has been widely recognized as a holistic mechanism for evaluating whether a collaborator possesses the required capabilities and resources to undertake specific tasks. Existing trust evaluation methods typically collect collaborators' historical performance data--such as task success rates and packet loss ratios--and aggregate these multi-dimensional metrics into a single scalar trust score using predefined weighting schemes~\cite{11296817}. However, scalar aggregation obscures dimension-specific deficiencies, making it difficult to identify which aspect of a collaborator is unreliable.

As future collaborative systems support increasingly diverse tasks, the requirements imposed on collaborators exhibit inherently multi-dimensional characteristics. For instance, safety-critical tasks require strictly high task completion accuracy and stable computational behaviors, whereas data-intensive tasks heavily prioritize storage capabilities and bandwidth performance~\cite{8080202}. Therefore, it is imperative for trust evaluation to transition from a traditional scalar paradigm to a multi-dimensional paradigm. The evaluation mechanism must independently quantify a collaborator's trustworthiness across each required dimension, producing a trust vector that directly corresponds to the task's multi-dimensional requirements. This approach not only yields a comprehensive and fine-grained profiling of the collaborator's trustworthiness but also guarantees that the selected collaborator satisfies the task requirements across all dimensions.
However, realizing multi-dimensional trust evaluation in collaborative systems poses several challenges.

\textit{How to effectively synchronize heterogeneous data for multi-dimensional trust evaluation?} In distributed collaborative systems, historical trust-related data collected from past collaborations is inevitably dispersed across different participants. To achieve accurate trust evaluation, the records of task outcomes from task owners and the corresponding states and resource conditions of participating collaborators have to be gathered for aggregated analysis. Due to clock drift across devices and varying delays introduced by network transmission, distributed data received by the trust evaluation server is often desynchronized and out of chronological order. Without effective synchronization, data from different collaborations may be mistakenly mixed together, and the data order within the same collaboration may be misaligned, ultimately leading to unreliable trust evaluation. While commonly used clock synchronization protocols (such as Network Time Protocol or Precision Time Protocol) can achieve time alignment across devices, they could incur substantial network communication overhead for distributed systems~\cite{7509657}. In fact, synchronizing all devices to a global absolute time coordinate system is unnecessary in trust evaluation. This is because the essence of trust evaluation is to discover the evolutionary patterns of collaborators' historical behaviors from the data collected during past collaborations. Therefore, we could consider some lightweight strategies, such as the monotonicity of local clocks within each data source and task identifiers, to align multi-dimensional data across collaborations.

\textit{How to jointly capture the intra-dimensional temporal trust evolution and the cross-dimensional dependencies?} 
A collaborator’s trustworthiness is inherently time-varying, and each trust dimension may exhibit complex evolution patterns. For instance, a collaborator’s task completion accuracy may gradually degrade due to device aging, abruptly deteriorate after a period of high trust because of strategic misbehavior, or intermittently fluctuate under unstable resource conditions~\cite{9629245}. Traditional trust estimation methods, such as weighted moving averages, primarily reflect recent statistical trends and lack the ability to actively identify such complex temporal patterns, particularly abrupt changes~\cite{10148640}. Furthermore, multiple trust dimensions are inherently interdependent and exhibit dynamic correlations. For example, anomalies in CPU stability often coincide with declines in task completion accuracy, while resource contention may simultaneously degrade timeliness and disrupt memory behavior. Existing trust evaluation approaches typically employ sequence models, e.g., standard Transformer-based architectures, to capture the evolution of trust. However, these models are primarily designed to model temporal dependencies. In multi-dimensional trust, an effective trust model must possess dual modeling capabilities: it should learn the temporal evolution of each dimension along the time axis while simultaneously capturing the dynamic coupling across dimensions, enabling more accurate and comprehensive trust assessment.

To address the aforementioned challenges, we propose TrustFormer, a Transformer-based multi-dimensional trust evaluation framework. TrustFormer integrates lightweight data synchronization with cross-temporal and cross-dimensional trust modeling to align heterogeneous trust-related data and evaluate collaborator trustworthiness according to task-specific multi-dimensional requirements. The contributions of this paper are summarized as follows.
\begin{itemize}
    \item We propose a task-specific multi-dimensional trust evaluation paradigm that provides a more comprehensive and realistic assessment of collaborator trustworthiness with respect to task requirements.

    \item We propose a lightweight data synchronization mechanism for multi-dimensional trust-related data by leveraging task identifiers and the monotonicity of local device clocks.

    \item  We design cross-temporal and cross-dimensional attention mechanisms that capture both the temporal evolution within each trust dimension and the interdependencies across different dimensions, enabling more effective modeling of multi-dimensional trust. 
\end{itemize}

\section{System Model and Problem Formulation}
\label{sec:system_model}

We consider a collaborative computing system comprising a trust server and a set of devices $\mathcal{A} = \{a_1, a_2, \ldots, a_I\}$. Within this system, each device $a_i \in \mathcal{A}$ operates in a dual capacity: it can act as a task owner that generates and offloads computational tasks, or as a collaborator that provisions its resources to execute tasks delegated by others. The trust server acts as a centralized coordinator responsible for aggregating the historical performance records of all collaborators. Upon the arrival of a new task request from a task owner, the trust server performs trust evaluation and selects the optimal collaborator. We assume the trust server is a fully credible entity equipped with sufficient computational capabilities.

\subsection{Task Model with Multi-Dimensional Requirements}
When device $a_i$ acts as a task owner, it generates a task $\eta_{a_i}$ characterized by
\begin{align}
    \eta_{a_i} = (\mathbf{q}^{\text{his}}, \mathbf{q}^{\text{res}}),
\end{align}
where $\mathbf{q}^{\text{his}} = [q_d^{\text{his}}]_{d \in \mathcal{D}^{\text{his}}} \in [0,1]^{|\mathcal{D}^{\text{his}}|}$ specifies the minimum trust requirements for potential collaborators across historical performance dimensions, and $\mathcal{D}^{\text{his}}$ denotes the set of such dimensions. In this paper, we consider $\mathcal{D}^{\text{his}}$ as $\{\mathrm{cu}, \mathrm{mu}, \mathrm{acc}, \mathrm{tli}\}$, where $\mathrm{cu}$ and $\mathrm{mu}$ represent CPU and memory stability, while $\mathrm{acc}$ and $\mathrm{tli}$ denote task completion accuracy and timeliness, respectively. Similarly, $\textbf{q}^{\text{res}} = [q^{\text{res}}_d]_{d\in\mathcal{D}^{\text{res}}} \in \mathbb{R}_{> 0}^{|\mathcal{D}^{\text{res}}|}$ specifies the requirements for potential collaborators' resources, where $\mathcal{D}^{\text{res}}$ is the set of physical resource dimensions.
In this paper, we assume that $\mathcal{D}^{\text{res}}$ consists of three dimensions: available CPU, storage, and throughput. Each $q^{\text{res}}_d$ dictates the lowest acceptable resource provision in its respective physical unit (e.g., GHz for CPU, GB for storage, or Mbps for throughput). This multi-dimensional requirement structure reflects the fact that different tasks impose distinct requirements on collaborator reliability and resource capabilities. It is worth noting that the number of requirement dimensions can be extended as needed.

\subsection{Task-Specific Multi-Dimensional Trust Model}
Upon receiving the task $\eta_{a_i}$ from task owner $a_i$, the trust server initiates trust evaluations for a set of candidate collaborators. Given the multi-dimensional nature of task requirements, relying on a conventional scalar trust value is insufficient. Therefore, it is imperative to evaluate potential collaborators across multiple dimensions to ensure that the selected collaborator can guarantee effective task completion. Specifically, for a candidate collaborator $a_j$, its comprehensive trust profile is quantified as the degree of credibility in satisfying the multi-dimensional requirements of task $\eta_{a_i}$ across both the historical performance dimensions $\mathcal{D}^{\text{his}}$ and the resource dimensions $\mathcal{D}^{\text{res}}$. Formally, the multi-dimensional trust of collaborator $a_j$ with respect to task $\eta_{a_i}$ is formulated as
\begin{align}
    \mathbf{T}(a_j) &= \left[ \mathbf{T}^{\text{his}}(a_j) \,;\, \mathbf{T}^{\text{res}}(a_j) \right] \in \mathbb{R}^{|\mathcal{D}^{\text{his}}| + |\mathcal{D}^{\text{res}}|}, \label{eq:T_vector}
\end{align}
where $\mathbf{T}^{\text{his}}(a_j) = [T_{d}^{\text{his}}]_{d \in \mathcal{D}^{\text{his}}} \in [0,1]^{|\mathcal{D}^{\text{his}}|}$ represents the multi-dimensional historical trust of collaborator $a_j$ evaluated from its historical performance records. Similarly, $\mathbf{T}^{\text{res}}(a_j) = [T_{d}^{\text{res}}]_{d \in \mathcal{D}^{\text{res}}} \in \{0,1\}^{|\mathcal{D}^{\text{res}}|}$ represents the multi-dimensional resource trust of collaborator $a_j$, indicating whether it possesses sufficient physical resources to satisfy the resource requirements of task $\eta_{a_i}$. Without loss of generality, we adopt a binary formulation, where $T_{d}^{\text{res}} = 1$ if the resource capacity of collaborator $a_j$ in dimension $d$ meets the corresponding task requirement; otherwise, $T_{d}^{\text{res}} = 0$. Based on the trust evaluation results, the trust server selects a suitable collaborator to execute the task $\eta_{a_i}$.

\subsection{Problem Formulation}

Given a task $\eta_{a_i} = (\textbf{q}^{\text{his}}, \textbf{q}^{\text{res}})$ submitted by task owner $a_i$, the trust server aims to select the optimal collaborator $a^*$ that maximizes the aggregate trust value subject to the task's multi-dimensional requirements. To this end, we formulate the collaborator selection as the following optimization problem:
\begin{align}
    a^* = \arg\max_{a_j \in \mathcal{A} \setminus \{a_i\}} \; & \left(\mathbf{1}^\top \mathbf{T}(a_j) \right), \label{eq:objective} \\
    \text{s.t.} \quad & T_{d}^{\text{his}} \geq q_d^{\text{his}}, \quad \forall d \in \mathcal{D}^{\text{his}}, \label{eq:c1} \\
     & T_{d}^{\text{res}} = 1, \quad \forall d \in \mathcal{D}^{\text{res}}, \label{eq:c_res}
\end{align}
where $\mathbf{1} \in \mathbb{R}^{|\mathcal{D}^{\text{his}}| + |\mathcal{D}^{\text{res}}|}$ is the all-ones vector, corresponding to uniform weighting across trust dimensions. The binary resource-trust dimensions are retained for a complete task-specific trust representation, while Constraint~\eqref{eq:c_res} restricts feasible candidates to those satisfying all resource requirements. Constraint~\eqref{eq:c1} ensures that each historical trust dimension meets the corresponding task-specified requirement. Task-specific weights can be incorporated by replacing $\mathbf{1}$ with a task-defined weight vector.

\section{TrustFormer for Multi-Dimensional Trust Evaluation}

This section introduces the proposed TrustFormer for multi-dimensional trust evaluation. TrustFormer incorporates lightweight synchronization strategies to accurately align the historical performance data of collaborators. By leveraging cross-temporal and cross-dimensional attention mechanisms, the framework effectively captures both the temporal dynamics within individual trust dimensions and the complex interdependencies across them. Consequently, TrustFormer facilitates a more comprehensive and accurate modeling of multi-dimensional trust.

\subsection{Multi-Dimensional Historical Data Collection}
\label{sec:data_sync}

Evaluating the multi-dimensional trust of collaborators requires collecting their performance data from historical collaborations across multiple metrics. This data should be obtained from both participants in each collaboration: task owners can only observe task outcomes (e.g., task completion accuracy and timeliness), whereas collaborators can report their internal runtime metrics (e.g., CPU utilization). We assume that both collaborators and task owners report data honestly; however, even if collaborators behave dishonestly, their performance can still be inferred from the task outcomes reported by task owners. The trust server coordinates the entire process and serves as the central repository for all performance records. We illustrate the data collection process through a single historical collaboration. Suppose the trust server establishes a collaboration between a task owner $a_m$ and a collaborator $a_o$ to execute a task $\eta_{a_m}$. Prior to task initiation, the trust server generates a globally unique cryptographic nonce, denoted as $\mathrm{ID}_{\eta_{a_m}}$, and records a server-side timestamp $t^{\text{start}}_{\eta_{a_m}}$ to mark the start of the collaboration. This $\mathrm{ID}_{\eta_{a_m}}$ is distributed to both parties, serving as the binding key that associates all subsequent data streams with this specific collaboration.

\textbf{Collaborator-side collection}: 
Throughout task execution, collaborator $a_o$ periodically samples and reports its local resource states to the trust server. According to the system model, historical CPU and memory utilization are required, defining the sampled dimension subset as $\{\mathrm{cu}, \mathrm{mu}\} \subset \mathcal{D}^{\text{his}}$. Each sampled record is represented as $(\text{ID}_{\eta_{a_m}}, t^{(n)}_{a_o}, x^{(n)}_{\mathrm{cu}}, x^{(n)}_{\mathrm{mu}})$, where $t^{(n)}_{a_o}$ is the device's local timestamp at the $n$-th sampling point, and $x^{(n)}_{\mathrm{cu}}, x^{(n)}_{\mathrm{mu}}$ are the measured CPU and memory utilization, respectively. The fixed-interval sampling (e.g., every 1 second) continues from task start until completion, producing a comprehensive sequence of records for each collaboration.

\textbf{Task-owner-side collection}:
Upon receiving the execution results, task owner $a_m$ independently verifies the outcomes and subsequently reports collaborator $a_o$'s performance to the trust server. This report covers two historical dimensions, $\{\mathrm{acc}, \mathrm{tli} \} \subset \mathcal{D}^{\text{his}}$, i.e., task completion accuracy and timeliness. Formally, this verification record is represented as $(\mathrm{ID}_{\eta_{a_m}}, t_{a_m}, x_{\mathrm{acc}}, x_{\mathrm{tli}}),$
where $t_{a_m}$ is task owner $a_m$'s local monotonic timestamp at the moment of verification.  $x_{\mathrm{acc}} \in [0,1]$ quantifies the validated task completion accuracy, while $x_{\mathrm{tli}} \in \{0, 1\}$ serves as a binary timeliness indicator, yielding $1$ if the task is completed within the prescribed deadline and $0$ otherwise. Once the verification record is received, the trust server records a local timestamp $t^{\text{end}}_{\eta_{a_m}}$ to define the server-side boundary for the completion of the task $\eta_{a_m}$. Any subsequent records tagged with $\mathrm{ID}_{\eta_{a_m}}$ that arrive beyond this boundary are discarded.

\subsection{Multi-Dimensional Historical Data Synchronization}
The trust server cannot utilize raw performance records directly, as variable network delays scramble their arrival order, and the embedded local timestamps originate from unsynchronized device clocks. In this study, strictly synchronizing all data to a unified global clock is not required. Instead, it is sufficient to accurately group the received data into their respective collaboration sessions and ensure that the chronological order within each individual performance dimension is correctly preserved. To reconstruct records from historical collaborations, the trust server first leverages $\mathrm{ID}_{\eta_{a_m}}$ to group all scattered records into their corresponding collaboration sessions. Subsequently, the trust server reconstructs the true chronological sequences of CPU and memory utilization, $\{x^{(n)}_{\text{cu}}\}_{n=1}^N$ and $\{x^{(n)}_{\text{mu}}\}_{n=1}^N$, by sorting the raw records according to their embedded local timestamps ($t^{(1)}_{a_o} < \cdots < t^{(N)}_{a_o}$). This localized temporal alignment is effective because records originating from the same physical device inherently guarantee the correct relative execution order.

To achieve a uniform multi-dimensional representation, the trust server aggregates the chronological sequences of the $\mathrm{cu}$ and $\mathrm{mu}$ dimensions into individual scalar values for collaborator $a_o$. Rather than using simple summary statistics that discard the inherent temporal structure, we compute the root mean square of successive differences in the ordered monitoring sequence to serve as a stability metric. The CPU stability dimension is calculated as
\begin{align}
    \widetilde{x}_{\mathrm{cu}} = \sqrt{\frac{1}{N-1} \sum_{n=1}^{N-1} \left( x^{(n+1)}_{\mathrm{cu}} - x^{(n)}_{\mathrm{cu}} \right)^2}.
\end{align}
The memory stability dimension, $\widetilde{x}_{\mathrm{mu}}$, is calculated identically using its respective sequence. A low value indicates stable resource consumption consistent with normal execution, whereas a high value suggests erratic behavior that may signal resource contention or abnormal execution. The correctness of these metrics critically depends on the ordering established in the synchronization step. Therefore, the multi-dimensional historical performance of collaborator $a_o$ in task $\eta_{a_m}$ is represented as a vector $\mathbf{X}_{\eta_{a_m}} = \left[ x_{\mathrm{acc}},\ x_{\mathrm{tli}}, \widetilde{x}_{\mathrm{cu}}, \widetilde{x}_{\mathrm{mu}}\right]\in \mathbb{R}^{1 \times |\mathcal{D}^{\text{his}}|}$, $|\mathcal{D}^{\text{his}}|=4$,
associated with the server-side completion timestamp $t^{\text{end}}_{\eta_{a_m}}$. 

\begin{figure*}[t!]
\centering
\includegraphics[scale=0.97]{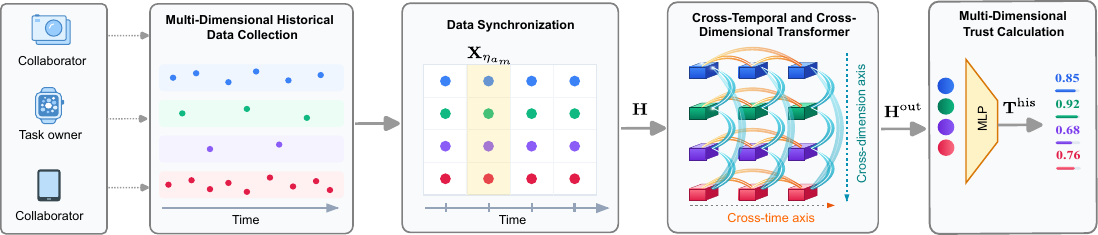}
\caption{TrustFormer leverages lightweight data synchronization and cross-temporal and cross-dimensional transformer to enable task-specific multi-dimensional trust evaluation and collaborator selection.}
\label{workflow}
\end{figure*}

\subsection{Data Preprocessing when New Tasks Arrive}
When a new task request ($\eta_{a_i}$) arrives from task owner $a_i$, the trust server initiates the historical trust evaluation for all potential collaborators. For collaborator $a_o$, the trust server retrieves the historical performance vectors of its $K$ previously completed collaborations. By vertically stacking these historical performance vectors in strictly chronological order, the server constructs the temporal multi-dimensional performance matrix $\mathbf{X}_{a_o}$ and its associated timestamp vector $\mathbf{t}_{a_o}$ as follows:
\begin{align}
\mathbf{X}_{a_o} &= [\dots; \mathbf{X}_{\eta_{a_m}}; \dots] \in \mathbb{R}^{K \times |\mathcal{D}^{\text{his}}|}, \\
\mathbf{t}_{a_o} &= [\dots, t^{\text{end}}_{\eta_{a_m}}, \dots]^{\top} \in \mathbb{R}^{K}, 
\end{align}
where the semicolon $(;)$ denotes row-wise vertical concatenation, ensuring that the rows of $\mathbf{X}_{a_o}$ temporally align with the elements of the column vector $\mathbf{t}_{a_o}$. Each column in $\mathbf{X}_{a_o}$ traces the temporal evolution of a single performance dimension.

\subsection{Multi-Dimensional Trust Representation Learning via Cross-Temporal and Cross-Dimensional Attentions}
To capture the complex temporal dynamics and inter-dimensional dependencies inherent in collaborator $a_o$'s multi-dimensional historical performance, we employ cross-temporal and cross-dimensional attention mechanisms.

\textbf{Embedding}: To project the matrix $\mathbf{X}_{a_o} \in \mathbb{R}^{K \times |\mathcal{D}^{\text{his}}|}$ into a representation suitable for learning, each element $x_{k,d}$ in $\mathbf{X}_{a_o}$, representing the $d$-th dimension of the $k$-th historical record, is projected via a dimension-specific linear layer to yield $\mathbf{e}_{k,d} \in \mathbb{R}^{d_{\text{model}}}$. For the temporal modeling, let $t_k$ denote the $k$-th element of $\mathbf{t}_{a_o}$. The time intervals $\Delta t_k = t_k - t_{k-1}$ between consecutive collaborations are encoded following the method in~\cite{DAI20262356}, combining a learnable positional embedding with a continuous time interval mapping to capture both sequential order and temporal distance, producing $\mathbf{u}_k \in \mathbb{R}^{d_{\text{model}}}$. The final embedding for dimension $d$ at the $k$-th historical record is $\mathbf{h}_{k,d} = \mathbf{e}_{k,d} + \mathbf{u}_k$, where $\mathbf{u}_k$ is shared across all dimensions within the same historical record. This process results in an embedding $\mathbf{H} = [\mathbf{h}_{k,d}] \in \mathbb{R}^{K \times |\mathcal{D}^{\text{his}}| \times d_{\text{model}}}, 1\leq k \leq K, 1 \leq d \leq |\mathcal{D}^{\text{his}}|$.

\textbf{Cross-temporal attention}: This attention captures how each performance dimension evolves across the $K$ historical collaborations. For each dimension $d \in \mathcal{D}^{\text{his}}$ independently, its temporal embedding sequence $\mathbf{H}_{:,d} \in \mathbb{R}^{K \times d_{\text{model}}}$ is extracted from $\mathbf{H}$. Standard multi-head self-attention (MSA) is applied along the temporal axis as follows:
\begin{align}
    \mathbf{Q}_d = \mathbf{H}_{:,d}\mathbf{W}^Q, \, \mathbf{K}_d &= \mathbf{H}_{:,d}\mathbf{W}^K, \, \mathbf{V}_d = \mathbf{H}_{:,d}\mathbf{W}^V, \\
    \mathbf{H}^{\text{time}}_d &= \text{softmax}\left(\frac{\mathbf{Q}_d \mathbf{K}_d^\top}{\sqrt{d_{\text{model}}}}\right)\mathbf{V}_d,
\end{align}
where $\mathbf{W}^Q, \mathbf{W}^K, \mathbf{W}^V \in \mathbb{R}^{d_{\text{model}} \times d_{\text{model}}}$ are learnable projection matrices. The attention matrix $\mathbf{Q}_d \mathbf{K}_d^\top \in \mathbb{R}^{K \times K}$ captures pairwise dependencies between all historical collaborations within dimension $d$, allowing the model to identify temporal patterns such as gradual performance degradation or abrupt behavioral changes. The per-dimension attention outputs are reassembled into the full representation and stabilized via residual connections and layer normalization, given by
\vspace{-0.07 in}
\begin{align}
    \bar{\mathbf{H}} &= \text{LayerNorm}(\text{Stack}(\mathbf{H}^{\text{time}}_1, \ldots, \mathbf{H}^{\text{time}}_{|\mathcal{D}^{\text{his}}|}) + \mathbf{H}),\\
    \hat{\mathbf{H}} &= \text{LayerNorm}(\text{MLP}(\bar{\mathbf{H}}) + \bar{\mathbf{H}}).
    \end{align}

\textbf{Cross-dimensional attention}: This attention captures correlations among different performance dimensions within each collaboration. For each collaboration $k$ independently, the embedding $\hat{\mathbf{H}}_{k,:} \in \mathbb{R}^{|\mathcal{D}^{\text{his}}| \times d_{\text{model}}}$ is extracted, containing the updated representations of all dimensions at the $k$-th collaboration. MSA is applied across dimensions
\begin{align}
    \mathbf{Q}_k = \hat{\mathbf{H}}_{k,:}\hat{\mathbf{W}}^Q, \, \mathbf{K}_k &= \hat{\mathbf{H}}_{k,:}\hat{\mathbf{W}}^K, \, \mathbf{V}_k = \hat{\mathbf{H}}_{k,:}\hat{\mathbf{W}}^V, \\
    \mathbf{H}_k^{\text{dim}} &= \text{softmax}\left(\frac{\mathbf{Q}_k \mathbf{K}_k^\top}{\sqrt{d_{\text{model}}}}\right)\mathbf{V}_k,
\end{align}
where $\mathbf{Q}_k \mathbf{K}_k^\top \in \mathbb{R}^{|\mathcal{D}^{\text{his}}| \times |\mathcal{D}^{\text{his}}|}$ captures pairwise dependencies among all performance dimensions. Since $|\mathcal{D}^{\text{his}}|$ is small, we directly apply MSA without the router mechanism in Crossformer \cite{chen2024crossformer}. The output follows the same residual and normalization structure as follows:
\begin{align}
    \check{\mathbf{H}} &= \text{LayerNorm}(\text{Stack}(\mathbf{H}^{\text{dim}}_1, \ldots, \mathbf{H}^{\text{dim}}_{K}) + \hat{\mathbf{H}}), \\
    \widetilde{\mathbf{H}} &= \text{LayerNorm}(\text{MLP}(\check{\mathbf{H}}) + \check{\mathbf{H}}).
\end{align}
This attention enables the model to learn dynamic cross-dimension correlations. For instance, a decline in CPU stability often co-occurs with a drop in task completion accuracy. The cross-temporal and cross-dimensional attention stages constitute a single time-dimension attention block. By stacking $L$ such blocks, the final output can be obtained as $\mathbf{H}^{\text{out}} \in \mathbb{R}^{K \times |\mathcal{D}^{\text{his}}| \times d_{\text{model}}}$, which encodes both the temporal dynamics and inter-dimensional dependencies of the collaborator's historical performance.

\subsection{Historical Trust Calculation}

To compute the multi-dimensional historical trust, we perform average pooling along the temporal axis for each dimension $d$ to obtain the temporally fused representation $\bar{\mathbf{h}}_d \in \mathbb{R}^{d_{\text{model}}}$. It is then passed through an MLP as follows:
\begin{align}
    T^{\text{his}}_{d} = \sigma(\text{MLP}_d(\bar{\mathbf{h}}_d)), \quad \forall d \in \mathcal{D}^{\text{his}},
\end{align}
where $\sigma(\cdot)$ is a sigmoid activation. The multi-dimensional historical trust evaluation for collaborator $a_o$ is
\begin{align}
    \mathbf{T}^{\text{his}}(a_o) = [T^{\text{his}}_{\mathrm{acc}}, T^{\text{his}}_{\mathrm{tli}}, T^{\text{his}}_{\mathrm{cu}}, \; T^{\text{his}}_{\mathrm{mu}}] \in [0,1]^{|\mathcal{D}^{\text{his}}|}.
\end{align}

\section{Resource Evaluation and Collaborator Selection}

After acquiring the historical trust $\mathbf{T}^{\text{his}}(a_o)$ of collaborator $a_o$, its resource trust is further evaluated across multiple dimensions. Based on the specific resource requirements of task $\eta_{a_i}$, collaborator $a_o$ submits its resource profiles to the trust evaluation server. The server then performs a dimension-wise evaluation: for each required dimension, the resource trust value is set to 1 if the reported capability satisfies the task requirement, and 0 otherwise. This process yields the multi-dimensional resource trust $\mathbf{T}^{\text{res}}(a_o)$ for collaborator $a_o$. This evaluation procedure is applied identically to all potential collaborators. Finally, in accordance with Eqs.~(\ref{eq:objective})-(\ref{eq:c_res}), the optimal collaborator is selected. Specifically, the chosen collaborator must satisfy the minimum trust thresholds across all dimensions required by task $\eta_{a_i}$ while achieving the maximum aggregate trust value.

\section{Result Analysis}
To validate the effectiveness of the proposed model, we adopt an empirical data-driven device modeling approach. To obtain realistic device behavior profiles, we deploy two representative collaborative tasks, namely face recognition and virus scanning, on DELL 5200, DELL 5820, and DELL 7060. The default resource requirements of each task for collaborators are set to 2 GHz of available CPU, 1 GB of storage, and 150 Mbps of throughput. By controlling the runtime conditions of devices, we systematically construct five representative behavior modes and collect the corresponding performance data: (i) reliable mode, where devices operate under interference-free conditions; (ii) low-capability mode, where intensive background processes are continuously executed, forcing devices to remain under high resource utilization; (iii) degradation mode, where background workload is gradually increased during execution, leading to progressive performance degradation over time; (iv) intermittent anomaly mode, where interfering processes are randomly activated and deactivated, causing devices to switch unpredictably between normal and high-load states; (v) strategic cheating mode, where devices operate normally in the first phase and then introduce sustained interference in the second phase, resulting in a sharp performance drop after an initially favorable period. For each device-task-behavior combination, 100 collaboration instances are conducted. In each instance, we collect the collaborator's performance data. Through statistical analysis, we extract the parameters across all dimensions for each behavioral pattern, thereby formulating the data-driven device behavior models.

Based on the constructed behavior models, we further build a simulation platform using Python and the NS-3 simulator that scales the three physical devices into 200 virtual devices. Each collaborator is assigned a behavior mode, with proportions set to $60\%$ reliable, $10\%$ low-capability, $10\%$ degradation, $10\%$ intermittent anomaly, and $10\%$ strategic cheating. The inter-arrival time between collaborations is sampled from an exponential distribution to emulate non-uniform collaboration patterns, resulting in a total of 20,000 records. The dataset is split chronologically into training ($70\%$), validation ($10\%$), and test ($20\%$) sets. The hyperparameters are set as follows: $d_{\text{model}}$ is 256, the number of attention heads is 4, and the number of block layers is 2. The model is trained using the mean squared error loss.

\vspace{-0.06 in}
\subsection{Impact of Synchronization on Trust Evaluation Accuracy}

To validate the necessity of the proposed data synchronization mechanism, we evaluate TrustFormer's evaluation accuracy when the input data suffer from varying degrees of temporal disorder. During inference, we randomly shuffle 10\%, 30\%, and 50\% of the collaborator-side monitoring records within each collaboration. This setup simulates the practical scenario where data synchronization is partially or entirely absent. As shown in Table~\ref{tab:sync}, we evaluate the prediction accuracy using mean squared error (MSE) and mean absolute error (MAE) under each disorder level. Both metrics increase monotonically with the degree of disorder, with MSE rising by 66.3\% and MAE by 31.4\% under 50\% disorder. Even a modest 10\% disorder causes a 12.2\% increase in MSE, confirming that data synchronization is an essential prerequisite for reliable multi-dimensional trust evaluation.

\begin{table}[t]
\centering
\caption{Impact of data disorder on trust evaluation accuracy. Results are reported as mean $\pm$ std over five runs.}
\label{tab:sync}
\begin{tabular}{c|cc}
\hline
Disorder Level & MSE & MAE \\
\hline
Synchronized & \textbf{0.0181 $\pm$ 0.0008} & \textbf{0.0923 $\pm$ 0.0021} \\
10\% & 0.0203 $\pm$ 0.0009 & 0.0987 $\pm$ 0.0024 \\
30\% & 0.0248 $\pm$ 0.0012 & 0.1098 $\pm$ 0.0028 \\
50\% & 0.0301 $\pm$ 0.0014 & 0.1213 $\pm$ 0.0033\\
\hline
\end{tabular}
\end{table}

\subsection{Trust Evaluation Accuracy Across Behavior Types}

To evaluate TrustFormer's trust evaluation accuracy under diverse collaborator behaviors, we compare it against LSTM and a standard Transformer by reporting the evaluation error separately for each behavior type. As shown in Table~\ref{tab:behavior}, all three methods achieve comparable trust evaluation accuracy for reliable collaborators, whose behavior is stationary and easily predictable. However, the performance gap widens significantly for collaborators with complex behavioral dynamics. For strategic collaborators, whose trustworthiness abruptly shifts from high to low, TrustFormer achieves an MSE of 0.0268, reducing the evaluation error by 40.8\% over LSTM and 28.3\% over the standard Transformer. A similar advantage is observed for degradation collaborators, where TrustFormer reduces MSE by 37.5\% and 22.1\% compared to the two baselines, respectively. These results indicate that TrustFormer enables more accurate trust evaluation under complex behavioral dynamics by explicitly capturing both temporal evolution and cross-dimension correlations.

\begin{table}[t]
\centering
\caption{Trust evaluation accuracy across behavior types.}
\label{tab:behavior}
\begin{tabular}{l|c|c|c}
\hline
\multirow{2}{*}{Behavior Type} & LSTM & Transformer & TrustFormer \\
 & MSE/MAE & MSE/MAE & MSE/MAE \\
\hline
Reliable       & 0.0156/0.0951 & 0.0143/0.0908 & \textbf{0.0138/0.0891} \\
Low-capability & 0.0247/0.1134 & 0.0218/0.1063 & \textbf{0.0201/0.1018} \\
Degradation    & 0.0389/0.1327 & 0.0312/0.1178 & \textbf{0.0243/0.1036} \\
Intermittent   & 0.0341/0.1198 & 0.0287/0.1094 & \textbf{0.0229/0.0973} \\
Strategic      & 0.0453/0.1364 & 0.0374/0.1231 & \textbf{0.0268/0.1043} \\
\hline
\end{tabular}
\end{table}

\begin{figure}[t!]
\centering
\includegraphics[scale=0.57]{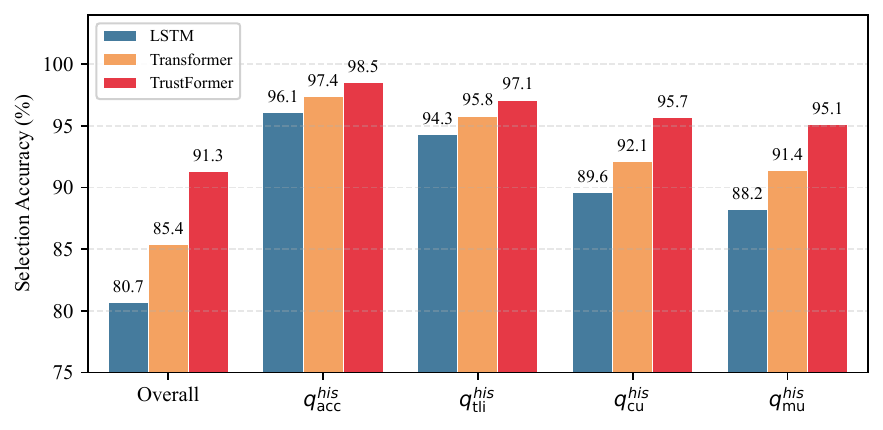}
\caption{The proposed TrustFormer achieves higher collaborator selection accuracy across all dimensions.}
\label{fig:selection}
\end{figure}

\subsection{Collaborator Selection Accuracy Comparison}
To evaluate the effectiveness of multi-dimensional trust evaluation for collaborator selection, we simulate 500 rounds of task assignment. In each round, a task is generated with a random requirement vector $\mathbf{q}^{\text{his}} = [q^{\text{his}}_{\mathrm{acc}}, q^{\text{his}}_{\mathrm{tli}}, q^{\text{his}}_{\mathrm{cu}}, q^{\text{his}}_{\mathrm{mu}}]$, where each component is uniformly sampled from $[0.5, 0.9]$. The resource requirements are kept fixed as described above. We report the collaborator selection accuracy at both the per-dimension level and the overall level, where the latter requires all dimensions to be simultaneously satisfied. 
As shown in Fig.~\ref{fig:selection}, TrustFormer consistently outperforms LSTM and Transformer across all dimensions. This gap is amplified in the overall accuracy, where TrustFormer achieves 91.3\% compared to 85.4\% (Transformer) and 80.7\% (LSTM), confirming that more accurate multi-dimensional trust evaluation translates directly into more reliable collaborator selection.

\section{Conclusion}
This paper has proposed TrustFormer, a task-specific multi-dimensional trust evaluation framework. TrustFormer achieves lightweight synchronization of multi-dimensional trust-related data. By employing cross-temporal attention and cross-dimensional attention, TrustFormer jointly captures the temporal evolution of each dimension and the dynamic correlations across dimensions from historical data, producing accurate multi-dimensional trust vectors to support collaborator selection. Experimental results have demonstrated that TrustFormer outperforms existing methods in both trust evaluation accuracy and collaborator selection reliability, with particularly notable advantages under complex behavioral patterns such as strategic deception and gradual degradation.

\vspace{-0.07 in}
\footnotesize

\end{document}